\documentclass[preprint]{revtex4}
\usepackage{enumerate}
\usepackage{graphicx}
\usepackage{dcolumn}
\usepackage{bm}
\usepackage{amsmath}
\usepackage{amsxtra}
\usepackage{amstext}
\usepackage{amssymb}
\usepackage{latexsym}
\begin{document}

%%%%%%%%%%%%%%%%%%%%%%%%TITLE%%%%%%%%%%%%%%%%%%%%%%%%
\title{
%\vskip-2.5truecm
%\rightline{\small{\tt ULB-TH/08-15}}
%\vskip2.5truecm
{ Relaxation of Chemical Reactions to Stationary States in the Chemical Affinities Space}
}
%%%%%%%%%%%%%%%%%%%%%%END_TITLE%%%%%%%%%%%%%%%%%%%%%%%

%%%%%%%%%%%%%%%%%%%%%%ADDRESSES%%%%%%%%%%%%%%%%%%%%%%
\author{ Giorgio SONNINO}
\email{gsonnino@ulb.ac.be}
\affiliation{
Universit{\'e} Libre de Bruxelles (U.L.B.)\\
Department of Theoretical Physics and Mathematics\\
Campus de la Plaine C.P. 231 - Bvd du Triomphe\\
B-1050 Brussels - Belgium\\
}

%%%%%%%%%%%%%%%%%%%%END_ADDRESSES%%%%%%%%%%%%%%%%%%%%%

%%%%%%%%%%%%%%%%%%%%%%ABSTRACT%%%%%%%%%%%%%%%%%%%%%%%
\begin{abstract}

Using the mass balance equations for chemical reactions, we show how the system relaxes towards a steady state in and out of the Onsager region. In the chemical affinities space, after fast transients, the relaxation process is a straight line when operating in the Onsager region, while out of this regime, the evolution of the system is such that the projections of the evolution equations for the forces and the shortest path on the flows coincide. 

\noindent For spatially-extended systems, similar results are valid for the evolution of the {\it thermodynamic mode} (i.e., the mode with wave-number ${\bf k} = {\bf 0}$).

\noindent These results allow us to obtain the expression for the affine connection of the space covered by the thermodynamic forces, close to the steady states. Through the affine connection, the nonlinear closure equations are derived.


\end{abstract}

\maketitle
%%%%%%%%%%%%%%%%%%%%%END_ABSTRACT%%%%%%%%%%%%%%%%%%%%%

%%%%%%%%%%%%%%%%%%%%%%TEXT_PAPER%%%%%%%%%%%%%%%%%%%%%%

\section{Introduction}
It is known that the evolution equations for the moments of the distribution functions we are confronted with the familiar difficulty appearing so often in statistical physics: an infinity hierarchy of moment equations. When we truncate the hierarchy, the resulting set of dynamical moments equations are not closed. Thus, a theory for the closure equations, (or the transport theory) should be formulated and combined with the truncated moment equations. A thermodynamical field theory (TFT) has been established addressing the closure problem for thermodynamic systems out of Onsager's region \cite{sonninog}. This theory allows to determine the nonlinear corrections to the Onsager theory so that the thermodynamic theorems for systems far from equilibrium are respected. Clearly, transport theory without a knowledge of microscopic dynamical laws can not be developed. Transport theory is only but an aspect of non-equilibrium statistical mechanics, which provides the link between micro and macro-levels of descriptions. This link appears indirectly in the "unperturbed" Onsager coefficients used as an input in the equations. The Onsager coefficients, which depend on the material under consideration, have to be calculated in the usual way by kinetic theory. 

\noindent The nonlinear contributions are derived starting from the De Donder-Prigogine principle [also refereed to as {\it the Thermodynamic Covariant Principle} (TCP)] stating that: "Thermodynamic systems are thermodynamically equivalent if, under transformation of fluxes and forces, the bilinear form of the entropy production remains unaltered". Flux-forces transformations leaving invariant the expression of the entropy production, are referred to as the {\it Thermodynamic Coordinate Transformations} (TCT). The TCP provides the possibility of defining an abstract space, {\it the thermodynamical forces space}, covered by the $r$ independent thermodynamic forces $X^i$, whose metric is identified with the symmetric part of the transport matrix. The law of evolution is not the dynamical law of particle motion, or the set i.e., of two-fluid macroscopic equations of plasma dynamics. It is rather the evolution of the thermodynamic forces of the system in the thermodynamical forces space. 

\noindent The nonlinear transport equations have been successively derived from two assumptions:
\begin{description}
\item {\bf 1.} The thermodynamic theorems valid when a generic thermodynamic system relaxes out of equilibrium are satisfied. In particular, the Glansdorff-Prigogine {\it Universal Criterion of Evolution} (UCE)] is fulfilled;
\item {\bf 2.} There exists a thermodynamic action, scalar under TCT, which is stationary for general variations in the transport coefficients and the affine connection of the thermodynamical forces space.
\end{description}
\noindent However, the determination of the thermodynamic action requires the knowledge of the affine connection. As we shall see, this expression can be deduced by analyzing several examples of relaxation. 

\noindent This theory has been successfully applied for studying transport processes in non equilibrium systems such as  magnetically confined plasmas, materials submitted to temperature and electric potential gradients or chemical reactions. In particular, in tokamak-plasmas, the thermodynamic forces and the conjugate flows are the generalized frictions and the Hermitian moments, respectively. Hence, we can determine the particle fluxes (electrons and ions) and energy losses as well as the (nonlinear) particle distribution functions. This has been accomplished in Ref. \cite{sonnino3} and in the paper reported in the footnote \footnote{G. Sonnino and P. Peeters, {\it Nonlinear transport processes in Tokamak Plasmas. Part II: The low-collisional regimes}, submitted to publication in the review {\it Physics of Plasmas} (2010).}.

\noindent In this paper, starting from the mass balance equations applied to chemical reaction-diffusion systems, we establish the following analytical results: 
\begin{description}
\item {($i$) After fast transients, in the Onsager region of the thermodynamical forces space, a homogeneous system relaxes towards a stable steady state along a straight line.} 
\item {($ii$) After fast transients, out of the Onsager region, a homogeneous system where $r$ chemical reactions take place simultaneously,  relaxes towards a stable steady state such that $\sum_{i=1}^rJ_i\ \mathcal{U}^i (X,\varrho) = \sum_{i=1}^rJ_i\ \mathcal{U}^i(X_{s.p.},\varrho)$ where $\mathcal{U}^i (X,\varrho)$ and $\mathcal{U}^i(X_{s.p.},\varrho)$ are the evolution equation for the chemical affinities and the shortest path in the chemical affinities space, respectively. $J_i$ denotes the vector of the chemical reaction rates. The trajectory traced out by the system and the shortest past are parametrized by $\varrho$. This parameter is defined in the Subsection \ref{subnl2} [see Eq.~(\ref{nlcr9b})].}
\end{description}
\noindent Similar theorems for the relaxation of the {\it thermodynamic mode} (i.e., the mode with wave-number ${\bf k} = {\bf 0}$) to the steady-state, can also be derived for spatially-extended systems.

\noindent These theorems allow us to obtain the expression for the affine connection of the space covered by the thermodynamic forces, close to the steady-states. Through the affine connection, the nonlinear transport relations are derived. 

\noindent The paper is organized as follows. In Section \ref{cher}, the simplest cases of isothermal and homogeneous chemical systems, relaxing in the Onsager region, are analyzed. A detailed study of chemical systems out of the linear region can be found in Section \ref{rts}. Since we are interested in the solution close to the stationary states, all expressions and matrices, entering in the calculations, are be developed in terms of a small parameter $\epsilon$ measuring the "{\it distance}" of the system from the steady state. We conclude in Section (\ref{concl}).

\noindent Such examples enable us to determine the expression of the affine connection for the thermodynamical forces space, close the non-equilibrium steady-states. In Ref. \cite{sonninog}, the nonlinear closure equations have been derived from assumption {\bf 2.} 

\vskip 0.2truecm
\section{Chemical Reactions in the Onsager Regime}\label{cher}
\vskip 0.2truecm
This section is devoted to the analysis of chemical systems relaxing towards a steady-state in Onsager's region of the thermodynamical forces space (i.e., the space of the chemical affinities) \cite{onsager}. For the sake of simplicity, we start by considering the relaxation of isothermal and homogeneous chemical reactions. We begin by studying in detail the relaxation of a triangular chemical system, maintained out of equilibrium by keeping constant the concentrations of two (among the three) chemical species. Our analysis is successively generalized by considering systems where $r$ independent chemical reactions take place simultaneously. Spatially-extended thermodynamic systems will be analyzed in the next section. 
\vskip 0.2truecm

\subsection{Unimolecular Triangular Chemical Reactions}\label{subl1}
\vskip 0.2truecm
\noindent Let us consider an ideal homogeneous system where the chemical reactions involve three substances $B$, $C$ and $D$ obeying to the following chemical scheme \cite{vidal}
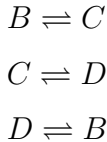
\begin{equation}\label{a1aa}
%\left\{
\!\!\!\!
\begin{array}{ll}
B\rightleftharpoons C& \\
C\rightleftharpoons D &\\
D\rightleftharpoons B &\nonumber
\end{array}
%\right.
\end{equation}
\noindent Indicating with $A_i$, $J_i$ and $C_i$ the {\it chemical affinities}, the {\it chemical reaction rate of reaction} $i$, and the concentrations of species $i$, respectively, we have 
\begin{equation}\label{a1a}
\qquad
\left\{ \begin{array}{ll}
A_1 = \mu_B-\mu_C & \ \ \mbox{;}\\
A_2 = \mu_C-\mu_D & \ \ \mbox{;}\\
A_3 = \mu_D-\mu_B & \ \ \mbox{;}
\end{array}
\right.
\quad
\left\{ \begin{array}{ll}
J_1=\overrightarrow {k}_1C_B-\overleftarrow{k}_1C_C \\
J_2=\overrightarrow {k}_2C_C-\overleftarrow{k}_2C_D \\
J_3=\overrightarrow {k}_3C_D-\overleftarrow{k}_3C_B 
\end{array}
\right.
\end{equation}
\noindent where $k_i$ are the direct ($\rightarrow$) and the inverse ($\leftarrow$) {\it kinetic constants} of the reactions and $\mu_i$ the {\it chemical potentials}. Note that, from Eqs~(\ref{a1a}) we find $A_1+A_2+A_3=0$, so only two affinities, say $A_1$ and $A_2$, are independent variables. At equilibrium, each chemical reaction rate vanishes, so the equilibrium concentrations ${\bar C}_{B,C,D}$ satisfy the conditions 
\begin{equation}
\quad
\left\{ \begin{array}{ll}
\overrightarrow {k}_1  {\bar C}_B=\overleftarrow{k}_1{\bar C}_C & \\
\overrightarrow {k}_2  {\bar C}_C=\overleftarrow{k}_2{\bar C}_D &\\
\overrightarrow {k}_3  {\bar C}_D=\overleftarrow{k}_3{\bar C}_B  & 
\end{array}
\right.
\end{equation}
\noindent We maintain the system out of equilibrium by keeping constant the concentrations of species $B$ and $C$:
\begin{equation}\label{a3aa}
\left\{ \begin{array}{ll}
C_B= {\bar C}_B =const.& \\
C_C = {\bar C}_C(1+\delta) = const.\ ;\ (\delta\ll1) &
\end{array}
\right.
\end{equation}
\noindent The dimensionless parameter $\delta$ measures the {\it distance} from equilibrium. This parameter is supposed to be small so that the linear analysis applies. Due to constraints Eqs~(\ref{a3aa}), the concentration of species $D$ varies. Its variation can be described by introducing a small dimensionless parameter $\varepsilon$
\begin{equation}\label{a3a}
C_D={\bar C}_D(1+\varepsilon)
\end{equation} 
\noindent The mass-balance equation for $C_D$ reads
\begin{equation}\label{a9}
\frac{{dC}_D}{dt}=J_2-J_3=\overrightarrow{k}_2C_C-\overleftarrow{k}_2C_D-\overrightarrow{k}_3C_D+\overleftarrow{k}_3C_B
\end{equation}
\noindent or, in terms of parameter $\varepsilon$
\begin{equation}\label{a3b}
\frac{d {\varepsilon}}{d t}= -(\overleftarrow{k}_2+\overrightarrow{k}_3){\varepsilon}+\overleftarrow{k}_2\delta
\end{equation}
\noindent This system admits the non equilibrium steady-state given by
\begin{equation}\label{a3c}
\varepsilon_{st.}=\frac{\overleftarrow{k}_2}{\overleftarrow{k}_2+\overrightarrow{k}_3}\delta
\end{equation}
\noindent Close to equilibrium, we can write
\begin{equation}\label{a4a}
\left\{ \begin{array}{ll}
J_1  \simeq\frac{\overleftarrow{k}_1\bar{C}_C}{R}\frac{A_1}{T} & \\
J_2  \simeq\frac{\overrightarrow{k}_2\bar{C}_C}{R}\frac{A_2}{T}  &\\
J_3  \simeq\frac{\overrightarrow{k}_3\bar{C}_D}{R}\frac{A_3}{T}  & 
\end{array}
\right.
\qquad{\rm and}\quad 
\left\{ \begin{array}{ll}
A_1=RT\ln\Bigl({\frac{{\bar C}_C{\bar C}_D}{C_CC_D}}\Bigr)\simeq RT(\delta - \varepsilon)  &\\
A_2=RT\ln\Bigl({\frac{C_DC_B}{{\bar C}_D{\bar C}_B}}\Bigr)\simeq  RT\varepsilon &\\
A_3=-(A_1+A_2) = -RT\delta&
\end{array}
\right.
\end{equation} 
\noindent where $R$ is the {\it gas constant} and $T$ is temperature of the system. By performing the linear transformations 
\begin{equation}\label{a5a}
\left\{ \begin{array}{ll}
A'_i=A_i& \\
J'_i = J_i-J_3 \qquad (i=1,2)&
\end{array}
\right.
\end{equation}
\noindent we obtain
\begin{eqnarray}\label{a6a}
&&J'_i=\sum_{j=1}^2L_{ij} X^j\\ \nonumber
&&\frac{1}{T}\sum_{i=1}^2J'_iA'_i=\frac{1}{T}\sum_{i=1}^3J_iA_i\qquad\bigl({\rm or,}\ \ \sigma'=\sigma\bigr)
\end{eqnarray}
\noindent where $\sigma$ indicates the {\it entropy production}. The expressions for the Onsager matrix $L_{ij}$ and the thermodynamic forces $X^i$ are
\begin{equation}\label{a7a}
L_{ij}=\frac{1}{R}
\begin{pmatrix}
\overleftarrow{k}_1\bar{C}_C+\overrightarrow{k}_3\bar{C}_D & \overrightarrow{k}_3\bar{C}_D\\
 \overrightarrow{k}_3\bar{C}_D& \overrightarrow{k}_2\bar{C}_C+\overrightarrow{k}_3\bar{C}_D
\end{pmatrix} \qquad {\rm ;}\quad
\left\{ \begin{array}{ll}
X^1\equiv\frac{A'_1}{T}=R(\delta-\varepsilon)& \\
X^2\equiv\frac{A'_2}{T}= R\varepsilon &
\end{array}
\right.
\end{equation}
\noindent Since the entropy production remains unaltered under transformations (\ref{a5a}), according to the general formulation, the systems $(A_i, J_i)$ and $(A'_i, J'_i)$ are thermodynamically equivalent \cite {sonninog}, \cite {dedonder}, \cite{prigogine_cov}. We can check that the thermodynamic forces satisfy the evolution equations
\begin{eqnarray}\label{a8a}
&&\frac{dX^i}{dt} =\sum_{j=1}^2A^i_j X^j +b^i\qquad (i=1,2)\qquad {\rm with}\nonumber\\
&& A^i_j \equiv-(\overleftarrow{k_2}+\overrightarrow{k_3})
\begin{pmatrix}
0 & 1\\
 1& 0
\end{pmatrix}\qquad {\rm ;}\quad
b^i\equiv R(\overleftarrow{k_2}+\overrightarrow{k_3})
\begin{pmatrix}
\varepsilon_{st.} \\
 \delta-\varepsilon_{st.}
\end{pmatrix}
\end{eqnarray}
\noindent The relaxation of the thermodynamic forces in terms of the {\it arc-parameter} $\varsigma$ is obtained. From the definition of $\varsigma$ 
\begin{equation}\label{a9a}
d\varsigma^2 = \sum_{i,j=1}^2L_{ij}dX^i dX^j
\end{equation}
\noindent we find 
\begin{equation}\label{a9aa}
\left\{ \begin{array}{ll}
X^1=R(\delta -\varepsilon_{st.})- L^{-1/2}(\varsigma -l) &\\
X^2 =R\ \varepsilon_{st.} + L^{-1/2}(\varsigma -l)
\end{array}
\right.
\end{equation}
\noindent where
\begin{equation}\label{a10a}
 L\equiv \frac{(\overrightarrow{k}_2+\overleftarrow{k}_3)}{R}{\bar C}_D\ \ ;\ \ l^2=\sum_{i,j=1}^2L_{ij}(X_0^i-X_{st.}^i)(X_0^j-X_{st.}^j)
\end{equation}
\noindent  $X_0^i$ are the initial conditions and $X_{st.}^i=[R(\delta -\varepsilon_{st.}),\  R\varepsilon_{st.}]$. As seen from Eq.~(\ref{a9aa}), in the Onsager region of the thermodynamic space, the system relaxes to the steady-state along a straight line.

\noindent Let us now quote the Glansdorff- Prigogine dissipative quantity $\mathcal{P}$. Noting that
\begin{equation}\label{a11a}
\varepsilon_{st.}-\varepsilon=\frac{L^{1/2}}{R}(l-\varsigma)
\end{equation}
\noindent after a little algebra, we have
\begin{eqnarray}\label{a12a}
&& P\equiv \sum_{i=1}^2J_i\frac{dX^i}{d\varsigma}=-RL^{-1/2}(\varepsilon_{st.}-\varepsilon)=
 -(l-\varsigma)\leq 0\nonumber\\
&& \mathcal{P} \equiv \sum_{i=1}^2J_i\frac{dX^i}{d t}= P \frac{d\varsigma}{d t}= - (\overrightarrow{k}_2+\overleftarrow{k}_3)(l-\varsigma)^2\propto - (l-\varsigma)^2 \leq 0
\end{eqnarray}
\noindent in agreement with the general results reported in Ref.~\cite{sonninog}. It is worthwhile to mention that the compact formalism provided by the differential geometry allows to obtain results (\ref{a12a}) directly, in only one line of calculation.
\vskip 0.4cm

\subsection{Several Independent Chemical Reactions}\label{subl2}

\noindent Let us now analyze the relaxation of $r$ coupled chemical reactions taking place simultaneously in a uniform system. In the Onsager region, the balance equations for the $n$ independent thermodynamic forces $X^\alpha$ read [see Eqs~(\ref{a8a})]
\begin{equation}\label{b1}
\frac{dX^\alpha}{dt} =\sum_{\alpha\beta}^rA^\alpha_\beta X^\beta +b^\alpha\qquad (\alpha,\beta=1,\dots,r)
\end{equation}
\noindent where $A^\alpha_\beta$ is a matrix independent of the thermodynamic forces (but it may depend on time) and $b^\alpha$ is a constant vector. It is useful to write the solutions of Eqs~(\ref{b1}) as
\begin{equation}\label{b2}
X^\alpha(t) = -\sum_{\beta=1}^r{\tilde A}^\alpha_{\beta st.}b^\beta + f^\alpha(t)\qquad{\rm with}\quad \lim_{t\rightarrow +\infty} f^\alpha(t) =0
\end{equation}
\noindent where ${\tilde A}^\alpha_{\beta st.}$ is the inverse of matrix $A^\alpha_\beta$ evaluated when $t\rightarrow+\infty$ and the steady-states, $X^\alpha_{st.}$, satisfy the condition
\begin{equation}\label{b1a}
X^\alpha_{st.}=-\sum_{\beta=1}^r{\tilde A}^\alpha_{\beta st.}b^\beta
\qquad{\rm with}\ \ {\tilde A}^\alpha_{\lambda st.}:\quad \sum_{\lambda=1}^r{\tilde A}^\alpha_{\lambda st.} A^\lambda_{\beta st.} =\delta^\alpha_\beta
\end{equation}
\noindent where $\delta^\alpha_\beta$ denoting the Kronecker delta. The expression of the arc-parameter reads
\begin{equation}\label{b3}
d\varsigma^2 = \sum_{\alpha,\beta=1}^rL_{\alpha\beta}dX^\alpha dX^\beta 
\ \Longrightarrow\ l-\varsigma =\int_t^{+\infty}\sqrt{\sum_{\alpha,\beta=1}^rL_{\alpha\beta}(t){\dot f}^\alpha(t){\dot f}^\beta(t)}\ dt = F(t)
\end{equation}
\noindent where "dot" indicates the time derivative. The second principle of thermodynamics ensures that $d\varsigma^2\geq 0$. Substituting Eq.~(\ref{b3}) into Eqs~(\ref{b2}), close to the steady-state, $\epsilon=(l-\varsigma)/l\ll 1$, we have
\begin{equation}\label{b4}
X^\alpha(\varsigma)= -\sum_{\beta=1}^r{\tilde A}^\alpha_{\beta st.}b^\beta + b_1^\alpha (\varsigma-l) +  b_2^\alpha (\varsigma-l)^2 +\cdots\qquad (b^\alpha_j=const.)
\end{equation}
\noindent  Now, taking into account  the symmetry
\begin{equation}\label{b5}
(\varsigma-l)\rightarrow -(\varsigma-l)
\qquad  \Longrightarrow\quad (X^\alpha-X^\alpha_{st.})\rightarrow -(X^\alpha-X^\alpha_{st.})
\end{equation}
\noindent we finally obtain
\begin{equation}\label{b6}
X^\alpha(\varsigma)= -\sum_{\beta=1}^r{\tilde A}^\alpha_{\beta st.}b^\beta + b_1^\alpha (\varsigma-l) + O(\epsilon^3)
\end{equation}
\noindent We may then summarize our results as follows: {\it In the Onsager region of the thermodynamical forces space, after fast transients, the thermodynamic system relaxes towards a stable steady state along a straight line}.
\vskip 0.2truecm
\section{Relaxation in the Nonlinear Region}\label{rts}
\vskip 0.2truecm
In the following sections we shall be concerned with the relaxation of chemical reactions towards a steady-state in the nonlinear region of the thermodynamical forces space i.e., out of Onsager's region. We first study the relaxation of isothermal and homogeneous chemical reactions and then we shall extend our analysis to spatially-distributed chemical systems. The study of these examples will enable us to establish the validity of the statements mentioned in the introduction. 

\subsection{Chemical Reactions in the Nonlinear Region}\label{subnl1}

\vskip 0.2truecm
In the first part of this section, we shall prove that, close to the steady-state, after fast transients, the Glansdorff-Prigogine dissipative quantity, computed through the balance equations coincides with the one related to the system relaxing along the shortest path in the chemical affinities space. In the second part, we shall show the validity of the statement ($ii)$ formulated in the Introduction. After fast transients, near the steady state, the projections of the evolution equation for the thermodynamic forces and the shortest path on the thermodynamic flows on the thermodynamic flows coincide. 

\noindent Let us consider a system consisting of $n$ components amongst which $r$ chemical reactions are possible. The (local) rate of change of the mass of component $\alpha$ is
\begin{equation}\label{nlcr1}
\frac{\partial\rho_\alpha}{\partial t} = -\nabla\cdot (\rho_\alpha{\bf v}_\alpha)+\sum_{i=1}^r\nu^i_\alpha J_i
\qquad (\alpha=1,2,\cdots, n)
\end{equation}
\noindent where $\rho_\alpha$ is the density (mass per unit volume) of $\alpha$, ${\bf v}_\alpha$ is the velocity of $\alpha$, and $\nu^i_\alpha J_i$ the production of $\alpha$ per unit volume in the $i^{\rm{th}}$ chemical reaction. The quantity $\nu^i_\alpha$ divided by the molecular mass $M_\alpha$ of component $k$ is proportional to the stoichiometric coefficient with which $\alpha$ appears in the chemical reaction $i$. The coefficients $\nu^i_\alpha$ are counted positive when components $\alpha$ appear in the second, negative when they appear in the first member of the reaction equations. Since the mass law is conserved in each separate chemical reaction we have 
\begin{equation}\label{nlcr2}
\sum_{\alpha=1}^n\nu^i_\alpha =0\qquad (i=1,2,\cdots, r)
\end{equation}
\noindent The closure equations for Eqs~(\ref{nlcr1}) read 
\begin{equation}\label{nlcr3}
J_i=\sum_{j=1}^r \tau_{ij}(A)A_j\qquad (i=1,2,\cdots, r)
\end{equation}
\noindent where we have introduced the chemical affinities $A_i$ of the reaction $i$ ($i=1,2,\cdots,r)$ and the transport coefficients $\tau_{ij}$ (which may depend on the chemical affinities). In general, $\tau_{ij}$ is an asymmetric matrix. The chemical affinities are linked to the {\it chemical potential} of component $\alpha$,  denoted by $\mu^k$, through the De Donder relation \cite{dedonder1}
\begin{equation}\label{nlcr4}
A_i=-\sum_{\alpha=1}^n\nu^i_\alpha \mu^\alpha\quad {\rm or}\quad X^i=-\frac{1}{T}\sum_{\alpha=1}^n\nu^i_\alpha \mu^\alpha \qquad (i=1,2,\cdots, r)
\end{equation}
\noindent where $X^i$ denote the thermodynamic force of the reaction $i$ ($X^i\equiv A_i/T$). We shall now analyze the relaxation towards stationary states. We shall firstly study the simpler case of isothermal and homogeneous chemical systems and, successively, we shall extend our treatment to space-dependent chemical systems.

\subsection{Isothermal and Homogeneous Chemical Reactions}\label{subnl2}

\noindent In this case the mass balance equations, Eqs~(\ref{nlcr1}), simplify to
\begin{equation}\label{nlcr5}
{\dot\rho}_\alpha= \sum_{i=1}^r\nu^i_\alpha J_i
\qquad (\alpha=1,2,\cdots, n)
\end{equation}
\noindent From Eq.~(\ref{nlcr3}-\ref{nlcr5}), we obtain the equation for the thermodynamic forces
\begin{eqnarray}\label{nlcr6}
&&{\dot X}^i= - \sum_{j=1}^rM^{ji}(X)J_j=-\sum_{j,l=1}^rM^{ji}(X)\tau_{jl}(X)X^l\qquad{\rm where}\\
&&M^{ji}(X)\equiv\sum_{\alpha,\beta=1}^n\nu^j_\beta\nu^i_\alpha\frac{\partial}{\partial\rho_\beta}\Bigl(\frac{\mu^\alpha}{T}\Bigr)
\qquad (i,j=1,2,\cdots,r)
\nonumber
\end{eqnarray}
\noindent By inverting matrix $M^{ij}(X)$ (see also the footnote \footnote{Matrix $M^{ij}$ is invertible only out of the steady-states. At the steady-state, we have $\dot{\rho}_\alpha=0$ or $\sum_{j=1}^r\nu^j_\alpha J_j(X_{st.})=0$ $(\alpha=1,2, \cdots, n)$ [see Eq.~(\ref{nlcr5})]. \label{inversion}}), the equations for the thermodynamic forces can be brought into the form
\begin{eqnarray}\label{nlcr7}
&&\!\!\!\!\!\!\!\! \!\!\!\!\!\!\!\! \!\!
J_i=-\sum_{j=1}^r{\tilde M}_{ij}(X){\dot X}^j=\sum_{j=1}^r\tau_{ij}(X)X^j   \qquad (i=1,2,\cdots,r)\quad ({\rm out\ of\ the\ steady\ state}) \\
&&\!\!\!\!\!\!\!\! \!\!\!\!\!\!\!\! \!\!\!\! 
\left\{ \begin{array}{ll}
\sum_{i=1}^r\nu^i_\alpha J_i(X_{st.})=\sum_{i,j=1}^r\nu^i_\alpha{\tilde M}_{ij}(X_{st.}){\dot X}^j\vert_{st.}=0\quad\ \ \!\ \!
 (\alpha=1,2,\cdots,n)&\\
\qquad\qquad\qquad\qquad\qquad\qquad\qquad\qquad\qquad\qquad\qquad\quad\!\!\!
 \qquad\quad\! ({\rm at\ the\ steady\ state})\nonumber \\
\sum_{i,j=1}^r{\tilde M}_{ij}(X_{st.}){\dot X}^i\vert_{st.}{\dot X}^j\vert_{st.}=0&
 \end{array}
\right.
\end{eqnarray}
\noindent where ${\tilde M}_{ij}(X)={\tilde M}_{ij}$ denotes the inverse of the matrix $M^{ij}$ i.e., $\sum_{l=1}^rM^{il}{\tilde M}_{jl}=\delta_j^i$.  The Glansdorff- Prigogine dissipative quantity $\mathcal{P}$ is immediately obtained from Eq.~(\ref{nlcr7})
\begin{equation}\label{nlcr8}
\mathcal{P}=  \sum_{i=1}^r J_i{\dot X}^i=-\sum_{i,j=1}^r{\tilde M}_{ij}(X){\dot X}^i{\dot X}^j\leq0
\end{equation}
\noindent Being ${\tilde M}_{ij}(X)$ a positive-definite matrix (for the demonstration see, for example, Ref.~\cite{degroot}), the dissipative quantity $\mathcal{P}$ is always negative throughout the evolution of the system. The inequality is only saturated at the steady-state. In literature, inequality Eq.~(\ref{nlcr8}) is referred to as the {\it Universal Criterion of Evolution} (UCE) \cite{prigogine1}.

\noindent In Ref.~\cite{sonninog} it is demonstrated the validity of the following theorem: {\it If a thermodynamic system relaxes towards a steady-state along the shortest path in the space of the thermodynamical forces then the UCE is automatically satisfied and}
\begin{equation}\label{nlcr9}
\mathcal{P}=  -{\dot\varrho}\int_{\varrho_0}^l\Bigl(\frac{d\varsigma}{d\varrho}\Bigr)^2d\varrho\leq 0
\end{equation}
\noindent where $\varsigma$ denotes the {\it arc-parameter}, defined as 
\begin{equation}\label{nlcr9a}
d\varsigma^2=\sum_{i,j=1}^r\tau_{ij}dX^idX^j\geq 0
\end{equation}
\noindent and $\varrho$ is a convenient parameter allowing to cast the equation for the shortest path in the space of the thermodynamic forces into the form
\begin{equation}\label{nlcr9b}
\frac{d^2X^i}{d\varrho^2}+\sum_{j,l=1}^r\Gamma^i_{jl}\frac{dX^j}{d\varrho}\frac{dX^l}{d\varrho}=0
\qquad (i=1,2,\cdots,r)
\end{equation}
\noindent Parameter $\varrho$ increases monotonically as the thermodynamic system evolves in time. It can also be chosen so that it vanishes when the thermodynamic system begins to evolve and it takes the (positive) value, say $l$, when the system reaches the steady-state \cite{sonninog}. $\Gamma^i_{jl}$ denotes the {\it affine connection} of the chemical affinities space \cite{sonninog}. For easy reference, the expression of $\Gamma^i_{jl}$ is reported in the Appendix of this paper. We want to check that, after fast transients, close to the steady state, Eq.~(\ref{nlcr8}) reduces indeed to the general expression, Eq.~(\ref{nlcr9}). To this purpose, we re-consider the forces equation, Eq.~(\ref{nlcr6}), the closure equation, Eq.~(\ref{nlcr3}), and constraint (\ref{nlcr2}). Defining $Y^i\equiv X^i-X^i_{st.}$, close to the steady-state $X_{st.}$, we may write
\begin{equation}\label{9b1}
M^{ji}(X)={\bar M}^{ji}(Y)+M^{ji}(X_{st.})\quad ;\quad J_i(X)={\bar J}_i(Y)+J_i(X_{st.})
\end{equation}
\noindent Taking into account that 
\begin{equation}\label{nlcr9c}
\sum_{j=1}^rM^{ji}(X_{st.})J_j(X_{st.})=0 \quad;\quad \sum_{j=1}^r{\bar M}^{ji}(Y)J_j(X_{st.})=0
\end{equation}
\noindent in terms of the variable $Y$, we obtain 
\begin{eqnarray}\label{nlcr9d}
&&
\left\{ \begin{array}{ll}
{\dot Y}^i= - \sum_{j=1}^r[{\bar M}^{ji}(Y)+M^{ji}(X_{st.})]{\bar J}_j(Y)\quad (i=1,2,\cdots,r) \qquad{\rm with}&\\
Y^i(+\infty)=0\qquad \bigl({\rm i.e.,}\ X^i(+\infty)=X^i_{st.}\bigr)&
\end{array}
\right.
 \\
&&{\bar J}_i(Y)=\sum_{j=1}^r {\bar\tau}_{ij}(Y)Y^j\quad ; \quad \sum_{\alpha=1}^n\nu^i_\alpha=0\qquad (i=1,2,\cdots,r)\nonumber 
\end{eqnarray}
\noindent We note that Eqs~(\ref{nlcr9d}) are manifestly invariant under transformations 
\begin{equation}\label{nlcr10}
\left\{ \begin{array}{ll}
t\rightarrow -t&\\
Y^i\rightarrow - Y^i&\\
\nu^i_\alpha\rightarrow -\nu^i_\alpha &
\end{array}
\right.
\qquad{\rm or}\qquad 
\left\{ \begin{array}{ll}
(\varrho-l)\rightarrow -(\varrho-l)&\\
Y^i\rightarrow - Y^i&\\
\nu^i_\alpha\rightarrow -\nu^i_\alpha &
\end{array}
\right.
\end{equation}
\noindent Recalling the definition of matrix $M^{ij}$ [see Eq.~(\ref{nlcr6})], we also find that, under Eqs~(\ref{nlcr10}), variables ${\bar J}_i$, ${\bar\tau}_{ij}$ and $M^{ij}$ transform as 
\begin{equation}\label{nlcr10b}
\!\!\ \left\{ \begin{array}{ll}
(\varrho-l)\rightarrow -(\varrho-l)&\\
Y^i\rightarrow - Y^i&\\
\nu^i_\alpha\rightarrow -\nu^i_\alpha &
\end{array}
\right.
\!\!\quad   \Longrightarrow\quad
\left\{ \begin{array}{ll}
{\bar J}_i(Y)\rightarrow -{\bar J}_i(Y)&\\
{\bar\tau}_{ij}(Y)\rightarrow {\bar\tau}_{ij}(Y)&\\
\Bigl[{\bar M}_{ij}(Y)+M_{ij}(X_{st})\Bigr]\rightarrow -\Bigl[{\bar M}_{ij}(Y)+M_{ij}(X_{st})\Bigr] &
\end{array}
\right.
\end{equation}
\noindent Hence, close to the steady-state [i.e., for $\epsilon=(\varrho-l)/l\ll1$], we may adopt the following expansions (see also the footnote \footnote {The matrix of the transport coefficients $\tau_{ij}(X)$ transforms, under the Thermodynamic Coordinate Transformations (TCT) [see Eqs~(\ref{nlcr19})],  as a thermodynamic tensor of the second order \cite{sonninog} i.e., 
${\bar\tau}_{ij}(Y)=\sum_{p,q=1}^r\frac{\partial X^p}{\partial Y^i}\frac{\partial X^q}{\partial Y^j}\tau_{pq}(X)$. Thus, under transformation $Y^i=X^i-X^i_{st.}$, we find ${\bar\tau}_{ij}(Y)=\tau_{ij}(X)=\tau_{ij}(Y+X_{st.})$.
\label{transport}})
\begin{equation}\label{nlcr12}
\left\{ \begin{array}{ll}
Y^i=X^i-X^i_{st.}=X^i_{(1)}(\varrho-l)+X^i_{(3)}(\varrho-l)^3+O(\epsilon^5) & \\
\frac{dY^i}{d\varrho}=\frac{dX^i}{d\varrho}=X^i_{(1)}+3X^i_{(3)}(\varrho-l)^2+O(\epsilon^4) & \\
{\bar \tau}_{ij}(Y)=\tau_{ij}(X)=\tau_{ij}(Y+X_{st.})=\tau_{ij}(X_{st.})+ \tau_{ij}^{(2)}(\varrho-l)^2+O(\epsilon^4) &\\
{\bar J}_i(Y)=J_i(X)-J_i(X_{st.})=J_i^{(1)}(\varrho-l)+J_i^{(3)}(\varrho-l)^3+O(\epsilon^5) & \\ 
{\bar M}^{ij}(Y)+M^{ij}(X_{st.})=M^{ij}(X)=M^{ij}_{(-1)}(\varrho-l)^{-1}+M^{ij}_{(1)}(\varrho-l)+M^{ij}_{(3)}(\varrho-l)^3+O(\epsilon^5)&\\
{\tilde M}_{ij}={\tilde M}_{ij}^{(1)}(\varrho-l)+{\tilde M}_{ij}^{(3)}(\varrho-l)^3+O(\epsilon^5)& 
\end{array}
\right.
\end{equation}
\noindent By inserting expansions Eqs (\ref{nlcr12}) into the mass balance equation, Eqs (\ref{nlcr9d}), at the first order we obtain
\begin{equation}\label{nlcr12a}
{\dot\varrho}X^i_{(1)}=-\sum_{j,p=1}^rM^{ji}_{(-1)}\tau_{jp}{(X_{st.})}X^p_{(1)}+O(\epsilon^2)
\end{equation}
\noindent or, by inverting the matrix $M^{ji}_{(-1)}$
\begin{equation}\label{nlcr13}
{\dot\varrho}\sum_{j=1}^r{\tilde M}_{ij}^{(1)}X^j_{(1)}=-\sum_{j=1}^r\tau_{ij}{(X_{st.})}X^j_{(1)}+O(\epsilon^2)
\end{equation}
\noindent Multiplying both sides of Eq.~(\ref{nlcr13}) by $(\varrho-l)X^i_{(1)}$ and summing over the index "$i$", we find 
\begin{eqnarray}\label{nlcr14}
{\dot\varrho}\sum_{i,j=1}^r\!{\tilde M}_{ij}^{(1)}(\varrho-l)X^i_{(1)}X^j_{(1)}=&& -\sum_{i,j=1}^r\!\tau_{ij}{(X_{st.})}X^i_{(1)}X^j_{(1)}(\varrho-l)+O(\epsilon^3)\nonumber\\
&&=\int_{\varrho_0}^l\sum_{i,j=1}^r\!\tau_{ij}{(X_{st.})}X^i_{(1)}X^j_{(1)}d\varrho+O(\epsilon^3)
\end{eqnarray}
\noindent However, from expansions Eqs~(\ref{nlcr12}), we have
\begin{equation}\label{nlcr15}
\left\{ \begin{array}{ll}
X^i_{(1)}=\frac{dX^i}{d\varrho}+O(\epsilon^2) & \\
\tau_{ij}{(X_{st.})}=\tau_{ij}+O(\epsilon^2) &\\
{\tilde M}_{ij}^{(1)}(\varrho-l)={\tilde M}_{ij}+O(\epsilon^3)& 
\end{array}
\right.
\end{equation}
\noindent Eq.~(\ref{nlcr14}) may then be rewritten as
\begin{equation}\label{nlcr16}
{\dot\varrho}\sum_{i,j=1}^r{\tilde M}_{ij}\frac{dX^i}{d\varrho}\frac{dX^j}{d\varrho}=\int_{{\varrho}_{0}}^l\Bigl(\sum_{i,j=1}^r\tau_{ij}\frac{dX^i}{d\varrho}\frac{dX^j}{d\varrho}\Bigr)d\varrho+O(\epsilon^3)=\int_{\varrho_0}^l\Bigl(\frac{d\varsigma}{d\varrho}\Bigr)^2d\varrho+O(\epsilon^3)
\end{equation}
\noindent where identity Eq.~(\ref{nlcr9a}) has been taken into account. Close to the steady-state, we finally obtain
\begin{equation}\label{nlcr17}
\mathcal{P}=\sum_{i=1}^rJ_i{\dot X^i}=-{\dot\varrho}^2\sum_{i,j=1}^r{\tilde M}_{ij}\frac{dX^i}{d\varrho}\frac{dX^j}{d\varrho}= -{\dot\varrho}\int_{\varrho_0}^l\Bigl(\frac{d\varsigma}{d\varrho}\Bigr)^2d\varrho+O(\epsilon^3)\leq 0
\end{equation}
\noindent in agreement with Eq.~(\ref{nlcr9}) and Ref.~\cite{sonninog}.

\noindent We are now interested in the evolution of the system near the steady-state. To this end, we reconsider the equation for the thermodynamic forces, Eq.~(\ref{nlcr6}), rewritten in terms of parameter $\varrho$
\begin{equation}\label{nlcr18}
{\dot\varrho}\frac{dX^i}{d\varrho}= - \sum_{j=1}^rM^{ji}(X)J_j
\qquad (i=1,2,\cdots,r)
\end{equation}
\noindent We note that this balance equation transforms in a covariant way under the following thermodynamic coordinate transformations (TCT) \cite{sonninog}
\begin{eqnarray}\label{nlcr19}
&&X'^i=\sum_{j=1}^r\frac{\partial X'^i}{\partial X^j} X^j\nonumber\\
&& J'_i=\sum_{j=1}^r\frac{\partial X^j}{\partial X'^i}J_j
\end{eqnarray}
\noindent Indeed, by inserting transformations (\ref{nlcr19}) into the Eq.~(\ref{nlcr18}) we find
\begin{eqnarray}\label{19a}
&&{\dot\varrho}\frac{dX'{}^i}{d\varrho}=\sum_{j=1}^r\Bigl(\sum_{l,q=1}^r\frac{\partial X'^{i}}{\partial X^l} \frac{\partial X'^j}{\partial X^q} M^{lq}\Bigr)J'_j=\sum_{j=1}^rM'{}^{ij}J'_j\qquad{\rm where}
\nonumber\\
&& M'{}^{ij}\equiv \sum_{l,q=1}^rM^{lq}\frac{\partial X'^{i}}{\partial X^l} \frac{\partial X'^j}{\partial X^q} 
\end{eqnarray}
\noindent Hence, matrix $M^{ij}$ transforms like a thermodynamic tensor of second rank. The TCT are the most general forces-transformations leaving invariant the expression of the entropy production and the Glansdorff-Prigogine dissipative quantity ${\mathcal P}$ \cite{sonninog}. ${\mathcal P}$ can be obtained by deriving both sides of Eq.~(\ref{nlcr18}) with respect to parameter $\varrho$. However, this operation should be performed in such a way to preserve the covariance under TCT.  For this, we undertake the {\it thermodynamic covariant differentiation along a curve} of both sides of Eq.~(\ref{nlcr18}) \cite{sonninog}, getting
\begin{equation}\label{nlcr20}
\frac{d^2X^i}{d\varrho^2}+\sum_{j,l=1}^r\Gamma_{jl}^i\frac{dX^j}{d\varrho}\frac{dX^l}{d\varrho}=-\sum_{j,l,p=1}^r\bigl( {\dot\varrho}^{-1}M^{ij}J_j\bigr)\!_{,l}\ {\dot\varrho}^{-1}M^{lp}J_p+{\dot\varrho}^{-2}
\sum_{j,l,p,q=1}^r\Gamma^i_{jl}M^{jp}M^{lq}J_pJ_q
\end{equation}
\noindent where comma $(,)$ stands for partial differentiation with respect to the thermodynamic forces. Eq.~(\ref{nlcr20}) may also be rewritten as 
\begin{eqnarray}\label{nlcr21}
&&\!\!\!\!\!\!\!\!\!\!\!\!\!\!\!\!{\mathcal U}^i(X,\varrho)\equiv \frac{d^2X^i}{d\varrho^2}+\sum_{j,l=1}^r\Gamma_{jl}^i\frac{dX^j}{d\varrho}\frac{dX^l}{d\varrho}+ h^i(X)={\mathcal U}^i(X_{s.path},\varrho)+ h^i(X)=0\quad {\rm where}\\
&&\!\!\!\!\!\!\!\!\!\!\!\!\!\!\!\!{\mathcal U}^i(X_{s.path},\varrho)\equiv \frac{d^2X^i}{d\varrho^2}+\sum_{j,l=1}^r\Gamma_{jl}^i\frac{dX^j}{d\varrho}\frac{dX^l}{d\varrho}\quad {\rm and}
\nonumber\\ 
&&\!\!\!\!\!\!\!\!\!\!\!\!\!\!\!\!h^i(X)\equiv \sum_{j,l,p=1}^r\bigl( {\dot\varrho}^{-1}M^{ij}J_j\bigr)\!_{,l}\ {\dot\varrho}^{-1}M^{lp}J_p-{\dot\varrho}^{-2}
\sum_{j,l,p,q=1}^r\Gamma^i_{jl}M^{jp}M^{lq}J_pJ_q\nonumber
\end{eqnarray}
\noindent  By using expansions (\ref{nlcr12}), close to the steady-state, we get (see Appendix)\begin{equation}\label{nlcr22}
h^i(X)\sim O(\epsilon)\qquad {\rm and} \qquad \sum_{i=1}^rJ_i h^i(X)\sim O(\epsilon^2)
\end{equation}
\noindent Near the steady-states, Eq.~(\ref{nlcr21}) satisfies the UCE. Indeed, multiplying Eq.~(\ref{nlcr21}) with the thermodynamic flows $J_i$ and summing, we obtain
\begin{eqnarray}\label{nlcr23}
\sum_{j=1}^rJ_i{\mathcal U}^i(X,\varrho)=&&\frac{d{\tilde P}}{d\varrho}
-\sum_{i,j=1}^r\tau_{ij}\frac{dX^i}{d\varrho}\frac{dX^j}{d\varrho}-\sum_{i,j,l=1}^r\tau_{ij,l}X^j
\frac{dX^i}{d\varrho}\frac{dX^l}{d\varrho}\nonumber\\
&&
+\sum_{i,j,l=1}^rJ_j\Gamma^j_{il}\frac{dX^i}{d\varrho}\frac{dX^l}{d\varrho}+O(\epsilon^2)=0
\end{eqnarray}
\noindent where ${\tilde P}=\sum_{i=1}^rJ_i \frac{dX^i}{d\varrho} $ and Eq.~(\ref{nlcr22}) has been taken into account. In the Appendix, it is shown the validity of the following identity
\begin{equation}\label{nlcr23a}
\sum_{i,j,l=1}^rJ_j\Gamma_{il}^j\frac{dX^i}{d\varrho}\frac{dX^l}{d\varrho}=\sum_{i,j,l=1}^r\tau_{ij,l}X^j\frac{dX^i}{d\varrho}\frac{dX^l}{d\varrho}
\end{equation}
\noindent From this equation and Eq.~(\ref{nlcr9a}), we obtain the expression
\noindent 
\begin{equation}\label{nlcr24}
\frac{d {\tilde P}}{d\varrho}-\Bigl(\frac{d\varsigma}{d\varrho}\Bigr)^2+O(\epsilon^2)=0
\end{equation}
\noindent Integrating from the initial condition to the steady-state, we get
\begin{equation}\label{nlcr26}
{\tilde P}(X^i_{st.})-{\tilde P}=\int_{\varrho_0}^l\Bigl(\frac{d\varsigma}{d\varrho}\Bigr)^2d\varrho+O(\epsilon^3)
\ge0
\end{equation}
\noindent However, ${\dot\varrho}{\tilde P}(X^i_{st.})=0$. Indeed 
\begin{eqnarray}\label{nlcr27}
{\dot\varrho}{\tilde P}(X^i_{st.})=&&{\dot\varrho}\sum_{i=1}^rJ_i(X_{st.})A^i_{(1)}=\sum_{i,j=1}^rJ_i(X_{st.})M^{ij}_{(-1)}J_j^{(1)}\nonumber\\
&&=\sum_{\alpha,\beta=1}^n\Bigl[\Bigl(\sum_{i=1}^r\nu^i_\alpha J_i(X_{st.})\Bigr)\Bigl(\sum_{j=1}^r\nu^j_\beta J_j^{(1)}\Bigr)\frac{\partial}{\partial\rho_\beta}\Bigl(\frac{\mu^\alpha}{T}\Bigr){\Big\arrowvert}_{(-1)}\Bigr]=0
\end{eqnarray}
\noindent where Eq.~(\ref{nlcr7}) has been taken into account. So, after fast transients, we finally obtain
\begin{equation}\label{nlcr28}
{\mathcal P}=\sum_{i=1}^rJ_i\frac{dX^i}{dt}={\dot\varrho}{\tilde P}=-{\dot\varrho}\int\Bigl(\frac{d\varsigma}{d\varrho}\Bigr)^2d\varrho\le0
\end{equation}
\noindent The inequality is only saturated at the steady-state. The main conclusion of our analysis is thus: {\it After fast transients, in the chemical affinities space, the system relaxes towards the steady-state so that}
\begin{eqnarray}\label{nlcr29}
&&\sum_{i=1}^rJ_i\mathcal{U}^i (X,\varrho) = \sum_{i=1}^rJ_i\mathcal{U}^i (X_{s.path},\varrho)+O(\epsilon^2)=0\qquad{\rm or}\nonumber\\
&&\sum_{i=1}^rJ_i\mathcal{U}^i (X,\varrho) - \sum_{i=1}^rJ_i\mathcal{U}^i (X_{s.path},\varrho)\sim O(\epsilon^2)
\end{eqnarray}
\noindent {\it Moreover, the Glansdorff-Prigogine dissipative quantity} ${\mathcal P}$, {\it differs from the one evaluated assuming that the system relaxes along the shortest path, by a factor of the order} $O(\epsilon^3)$. We have also verified that, during the relaxation, the UCE is satisfied. Note that in Onsager's region, $\varrho\propto\varsigma$ (see Ref. \cite{sonninog}) and $d\tau_{ij}/d\varsigma=0$. Hence, Eq.~(\ref{nlcr20}) reduces to
\begin{equation}\label{nlcr30}
\frac{d^2X^i}{d\varsigma^2}=0+O(\epsilon)\quad {\rm or} \quad 
X^i=X_{st.}^i+b_1^i (\varsigma -l)+O(\epsilon^3)
\end{equation}
\noindent In agreement with Eq.~(\ref{b6}).

\subsection{Spatially-extended chemical systems}\label{subnl3}

\noindent Previous results can easily be generalized for spacial chemical systems. In this section we shall show a general result valid for the relaxation of the thermodynamical mode (i.e., the mode with wave-number ${\bf k} = {\bf 0}$). For this we shall adopt the following strategy: firstly  we develop the space-time dependent thermodynamic forces (${\mathcal X}^i$), flows (${\mathcal J}_i$) and transport coefficients in (spatial) Fourier's series. Then, we perform the same calculations as in the homogeneous case by taking into account the {\it slaving principle} \cite{haken} (see the footnote \footnote{The slaving principle establishes that in a relaxation process, contributions from different wave-numbers are negligible with respect to those with same wave-numbers.\label{slaving}}). Close to the steady-state, we have \cite{sonninog}
\begin{eqnarray}\label{nlcr}
&&\!\!\!\!\!\!\!\!\!\!\!\!\!\!\!\!\!\!\!\!\!\!\!\!\!\!\!\int_V \sum_{j=1}^r{\mathcal J}_j\frac{d{\mathcal X}^j}{d\varrho}dv\simeq V \sum_{j=1}^r{\widehat J}_{({\bf 0})j}\frac{d{\widehat X}^j_{({\bf 0})}}{d\varrho}\leq 0
\qquad\qquad\quad {\rm with}\nonumber\\
&&\!\!\!\!\!\!\!\!\!\!\!\!\!\!\!\!\!\!\!\!\!\!\!\!\!\!\!
{\widehat J}_{({\bf k})j}(t)=\frac{1}{V}\int_V {\mathcal J}_j({\bf r},t)
\exp(-{\rm i}{\bf k}\cdot{\bf r})dv\nonumber\\
&&\!\!\!\!\!\!\!\!\!\!\!\!\!\!\!\!\!\!\!\!\!\!\!\!\!\!\!
{\widehat X}^j_{({\bf k})}(t)=\frac{1}{V}\int_V{\mathcal X}^j({\bf r},t)
\exp(-{\rm i}{\bf k}\cdot{\bf r})dv
\end{eqnarray}
\noindent $dv$ denotes a (spatial) volume element of the system, and the integration is over the entire space $V$ occupied by the system in question. Suffix $"({\bf 0})"$ denotes the {\it thermodynamic mode}. The evolution equations for $\rho_{({\bf k})\alpha}$ reads
\begin{eqnarray}\label{secr1}
&&{\dot{\widehat{\rho}}}_{({\bf k})\alpha}=-{\bf k}\cdot\bigl(\widehat{\rho}_{({\bf k})\alpha}\star{{\widehat{\bf v}}}_{({\bf k})}\bigr)+\sum_{j=1}^r\nu^j_\alpha {\widehat J}_{({\bf k})j}\qquad (\alpha=1,2,\dots,n)\\
&&{\widehat X}^j_{({\bf k})}=-\sum_{\alpha=1}^n\nu^i_\alpha\widehat{{\tilde\mu}}_{({\bf k)}}^\alpha\qquad (i=1,2,\dots,r)\nonumber
\end{eqnarray}
\noindent where symbol $"\star"$ indicates the convolution and ${\tilde\mu}^\alpha\equiv\mu^\alpha/T$. At the thermodynamic mode we have
\begin{eqnarray}\label{secr2}
&&{\dot{\widehat{\rho}}}_{({\bf 0})\alpha}=\sum_{j=1}^r\nu^j_\alpha {\widehat J}_{({\bf 0})j}+
{\rm O.T.}({\bf k},{\bf k'})_{{\bf k}\neq{\bf k'}\neq {\bf 0}}
\qquad (\alpha=1,2,\dots,n)\\
&&{\widehat X}^j_{({\bf 0})}=-\sum_{\alpha=1}^n\nu^i_\alpha\widehat{{\tilde\mu}}_{({\bf 0)}}^\alpha\qquad (i=1,2,\dots,r)\nonumber
\end{eqnarray}
\noindent where ${\rm O.T.}({\bf k},{\bf k'})_{{\bf k}\neq{\bf k'}\neq {\bf 0}}$ stands for {\it other contributions from different wave-numbers}. Due to the slaving principle, near the steady-state, the evolution equations for ${\widehat{\rho}}_{({\bf 0})\alpha}$ reduce to
\begin{eqnarray}\label{secr3}
&&{\dot{\widehat{\rho}}}_{({\bf 0})\alpha}=\sum_{j=1}^r\nu^j_\alpha {\widehat J}_{({\bf 0})j}
\qquad (\alpha=1,2,\dots,n)\\
&&{\widehat X}^j_{({\bf 0})}=-\sum_{\alpha=1}^n\nu^i_\alpha\widehat{{\tilde\mu}}_{({\bf 0)}}^\alpha\qquad (i=1,2,\dots,r)\nonumber
\end{eqnarray}
\noindent By performing now the same calculations as for the homogeneous case, we arrive to the following final result: {\it the thermodynamic mode} (${\bf k} = {\bf 0}$) {\it relaxes to the steady-state according to the law}
\begin{equation}\label{secr4}
\sum_{j=1}^r{\widehat J}_{({\bf 0})j}\mathcal{U}^j ({\widehat X}_{({\bf 0})},\varrho) = \sum_{j=1}^r{\widehat J}_{({\bf 0})j}\mathcal{U}^j ({\widehat X}_{({\bf 0})s.path},\varrho)=0+O(\epsilon^2)
\end{equation}
\vskip 0.2truecm
\section{Conclusions}\label{concl}
\vskip 0.2truecm

The Universal Criterion of Evolution theorem has been demonstrated by Glansdorff and Prigogine through the balance equations. From this theorem and by analyzing examples of relaxation of chemical systems close to steady-states, we may obtain the expression for the affine connection of the thermodynamical forces space. The construction of the correct affine connection is made "step by step".  At the first stage, we suppose for simplicity that the matrix of the transport coefficients is symmetric i.e., $\tau_{ij}=g_{ij}$. We multiply Eq.~(\ref{nlcr18}) by $g_{ip}$ and sum over the index $p$. By taking the derivative, with respect to parameter $\varrho$, of both sides of the resulting equation, we find
\begin{equation}\label{concl1}
\sum_{l=1}^rg_{il}\frac{d^2X^l}{d\varrho^2}+\frac{1}{2}\sum_{l,p=1}^r\bigl(g_{il,p}+g_{ip,l}\bigr)\frac{dX^l}{d\varrho}\frac{dX^p}{d\varrho}=\!\!\!\!
\sum_{j,l,p,q=1}^r\Bigl({\dot\varrho}^{-1}g_{il}M^{jl}J_j\Bigr)_{,p}{\dot\varrho}^{-1}M^{pq}J_q
\end{equation}
\noindent From Eq.~(\ref{a14}) (see the Appendix) we have 
\begin{equation}\label{concl2}
\sum_{j,l,p,q=1}^r\Bigl({\dot\varrho}^{-1}g_{il}M^{jl}J_j\Bigr)_{,p}{\dot\varrho}^{-1}M^{pq}J_q=-6\sum_{j=1}^rg_{ij}^{(0)}X^j_{(1)}(\varrho-l)+O(\epsilon^3)
\end{equation}
\noindent So, close to the steady-state, we obtain 
\begin{equation}\label{concl3}
\sum_{l=1}^rg_{il}\frac{d^2X^l}{d\varrho^2}+\frac{1}{2}\sum_{l,p=1}^r\bigl(g_{li,p}+g_{ip,l}\bigr)\frac{dX^l}{d\varrho}\frac{dX^p}{d\varrho}=O(\epsilon)
\end{equation}
\noindent At the lowest order, previous equation may be rewritten as
\begin{equation}\label{concl3a}
\frac{d^2X^i}{d\varrho^2}+\frac{1}{2}\sum_{j,l,p=1}^rg^{ij}\bigl(g_{lj,p}+g_{jp,l}\bigr)\frac{dX^l}{d\varrho}\frac{dX^p}{d\varrho}=O(\epsilon)
\end{equation}
\noindent where $g^{ip}$ denotes the inverse of the matrix $g_{jp}$ (i.e., $\sum_{p=1}^rg^{ip}g_{jp}=\delta^i_j$). It is not difficult to show that Eq.~(\ref{concl3a}) satisfies the UCE but it is {\it not} covariant under the TCT. This because the sum of the terms $\frac{1}{2}\sum_{l,p=1}^rg^{ij}\bigl(g_{jl,p}+g_{jp,l}\bigr)$, does not transform as an affine connection under the TCT. This condition is, however, satisfied by adding the Levi-Civita term $-1/2\sum_{j=1}^rg^{ij}g_{lp,j}$. Now, if one wants the Universal Criterion of Evolution satisfied also when the system relaxes along a shortest path, without imposing a priori {\it any} restrictions on transport coefficients, an extra term to the Levi-Civita affine connection should be added. It can be checked that the most general expression for this extra term is $1/(2\sigma)\sum_{j=1}^rX^iX^j g_{lp,j}$ \cite{sonninog}. Hence, the affine connection my be written as
\begin{equation}\label{concl4}
{\Gamma}^i_{lp}=\frac{1}{2}\sum_{j=1}^r\Bigl[g^{ij}\Bigl(\frac{\partial g_{lj}}{\partial X^p}+\frac{\partial g_{jp}}{\partial X^l}-\frac{\partial g_{lp}}{\partial X^j}\Bigr)\Bigr]+\frac{1}{2\sigma}\sum_{j=1}^rX^i X^j\frac{\partial g_{lp}}{\partial X^j}
\end{equation}
\noindent By performing the same calculations as in Section \ref{rts}, we can easily check that the evolution equation (parametrized with $\varrho$), obtained by taking the covariant derivative of both sides of Eq.~(\ref{nlcr18}) with affine connection (\ref{concl4}), is covariant under TCT and satisfies the UCE. In ref. \cite{sonninog} it is shown that in the general case, (i.e. when $\tau_{ij}=g_{ij}+f_{ij}$), the expression of the affine connection is given by Eq.~({\ref{a1}), which generalizes Eq.~(\ref{concl4}). The covariant derivative, made with the affine connection Eq.~(\ref{a1}) [or, with Eq.~(\ref{concl4}) when $\tau_{ij}$ is symmetric], is referred to as the {\it thermodynamic covariant derivative} \cite{sonninog}.

\noindent In the examples examined in this paper, the small parameter $\epsilon$ measures the distance of the system from the steady state. In the article cited in the footnote \footnote{G. Sonnino, {\it Relaxation of Magnetically Confined Tokamak-Plasmas to Mechanical Equilibria}, to be submitted for publication in the review {\it Physics of Plasmas} (2010).}, it is analyzed the relaxation of magnetically confined plasmas in a toroidal geometry. The characteristic feature of these equations is the presence of a small parameter, the {\it drift parameter}, defined as the Larmor radius over a macroscopic length \cite{balescu2}.  In this situation, the long-time behaviour of the solution, describing the evolution of the system near the steady state, may be obtained by using the multiple time-scale perturbation expansion (see, for example, the book cited in Ref. \cite{davidson}). The results obtained are valid generally because the dynamics include all relevant moment equations and the parameter $\epsilon$ is not related to the distance of the system from the stationary states. It is possible to show that, even in this case, the results $(i)$ and $(ii)$ (see the Introduction) remain valid. 

\noindent A non-Riemannian geometry may successively be constructed out of the components of the affine connections \cite{sonninog}, \cite{eisenhart}. 

\noindent The main conclusion of the present analysis is thus:

\noindent  {\it Close to steady-states, the geometry of the thermodynamic space is non-Riemannian with affine connection given by Eq.~(\ref{a1})}. The knowledge of the expression for the affine connection is a fundamental prerequisite for the construction of the (nonlinear) closure theory on transport processes. In Ref.~\cite{sonninog}, the curvature tensor and the nonlinear transport equations have successively been derived from Eq.~(\ref{a1}) and by introducing the following assumption:

\noindent {\it There exists a thermodynamic action, scalar under $TCT$, which is stationary with respect to arbitrary variations in the transport coefficients and the affine connection}. 

\noindent From this principle, a set of closure equations, constraints, and boundary conditions have been derived. These equations determine the nonlinear corrections to the linear ("Onsager") transport coefficients. The validity of the nonlinear transport equations has been largely tested by analyzing several symmetric transport processes, such as the thermoelectric effect and magnetically confined plasmas \cite{sonnino}. Transport processes in tokamak plasmas are of particular interest. In this case, even in absence of turbulence, the state of the plasma remains close to (but, it is not in) a state of local equilibrium. This prevents the transport relations from being linear. The nonlinear transport equations have been used for computing the particle and energy losses in tokamak-plasmas and for deriving the distribution density functions for the species $\alpha$ in several collisional regimes \cite{sonnino3}.

\vskip 0.2truecm
\section{Acknowledgments}
I would like to pay tribute to Prof. I.Prigogine. My strong interest in this domain of research is due to him, who promoted the {\it Brussels School of Thermodynamics} at the U.L.B., where I took my doctorate in Physics. I am grateful to Dr.~J.Evslin from the SISSA (International School for Advanced Studies), Prof.~M.Tlidi, from the Universit{\'e} Libre de Bruxelles (U.L.B.) and Dr.~G.Breyiannis, from the EURATOM/ENEA Italian Fusion Association in Frascati (Rome) for the useful discussions and suggestions. 
\vskip 0.2truecm
\appendix 
\noindent \section{The Expression of the Affine Connection and its Main Properties.}\label{appx}
\vskip 0.2truecm
In this section we shall prove the validity of Eqs~(\ref{nlcr22}) and Eq.~(\ref{nlcr23a}). The expression of the affine connection is reported in Ref. \cite{sonninog} 
\begin{eqnarray}\label{a1}
\!\!\!\!\!\!\!\!\!\!\!\!\!\!\!\Gamma^i_{jq'}=\!\!\!\!&&\sum_{p,q=1}^r{\check N}^{ip}g_{pq}\begin{Bmatrix} 
q \\ jq'
\end{Bmatrix}+\sum_{p,q,s=1}^r\Bigl[\frac{{\check N}^{ip}}{2\sigma}g_{ps}X^sX^q\frac{\partial g_{jq'}}{\partial X^q}
+\frac{{\check N}^{ip} }{2\sigma}g_{ps}X^s X^q\Bigl(\frac{\partial f_{jq}}{\partial X^{q'}}+\frac{\partial f_{q'q}}{\partial X^j}\Bigr)\nonumber\\
&&
\qquad\qquad\qquad\ \ \!
+\frac{{\check N}^{ip}}{2\sigma}f_{ps}X^s X^q\Bigl(\frac{\partial g_{jq}}{\partial X^{q'}}+\frac{\partial g_{q'q}}{\partial X^j}\Bigr)\Bigr]
\end{eqnarray}
\noindent where the thermodynamic Christoffel symbols of the second kind are introduced
\begin{equation}\label{a2}
\begin{Bmatrix} 
q \\ jq'
\end{Bmatrix}=\frac{1}{2}\sum_{p'=1}^r g^{qp'}\Bigl(\frac{\partial g_{p'j}}{\partial X^{q'}}+\frac{\partial g_{p'q'}}{\partial X^j}-\frac{\partial g_{j q'}}{\partial X^{p'}}\Bigr)
\end{equation}
\noindent and matrices ${\check N}^{ij}$, $g_{ij}$ and $f_{ij}$ are defined as
\begin{eqnarray}\label{a3}
&&\!\ \!N_{ij}\equiv g_{ij}+\frac{1}{\sigma}\sum_{p,q=1}^r\bigl(f_{ip}g_{jq}+f_{jp}g_{iq}\bigr)X^p X^q\quad{\rm with}\ \ \ {\check N^{il}}:\ \ \sum_{p=1}^r{\check N}^{ip}N_{jp}=\delta_{j}^{i}\\
&&g_{ij}\equiv\frac{1}{2}(\tau_{ij}+\tau_{ji})\qquad ;\qquad f_{ij}\equiv\frac{1}{2}(\tau_{ij}-\tau_{ji})\nonumber
\end{eqnarray}
\noindent $\sigma$ denotes the entropy production of the system
\begin{equation}\label{a4}
\sigma=\frac{1}{T^2}\sum_{i,j=1}^r\tau_{ij}A_iA_j=\sum_{i,j=1}^r\tau_{ij}X^iX^j
\end{equation}
\noindent We can check that matrices ${\check N}^{ij}$, $g_{ij}$ and $f_{ij}$ possess the following properties
\begin{equation}\label{a5}
\sigma=\sum_{i,j=1}^r\tau_{ij}X^iX^j=\sum_{i,j=1}^rg_{ij}X^iX^j\quad ;\quad \sum_{i,j=1}^rf_{ij}X^iX^j=0\quad ; \quad \sum_{p=1}^rJ_p{\check N}^{ip}=X^i
\end{equation}
\noindent Multiplying Eq.~(\ref{a1}) with the factor $J_i\ dX^j/d\varrho\ dX^p/d\varrho$ and summing over the indexes, we obtain Eq.~(\ref{nlcr23a})
\begin{equation}\label{a6}
\sum_{i,j,p=1}^rJ_i\Gamma^i_{jp}\frac{dX^j}{d\varrho}\frac{dX^p}{d\varrho}=\!\! 
\sum_{j,p,q=1}^rg_{jq,p}X^q\frac{dX^j}{d\varrho}\frac{dX^p}{d\varrho}+\!\!\sum_{j,p,q=1}^rf_{jq,p}X^q\frac{dX^j}{d\varrho}\frac{dX^p}{d\varrho}=\!\!\sum_{j,p,q=1}^rX^q\tau_{jq,p}\frac{dX^j}{d\varrho}\frac{dX^p}{d\varrho}
\end{equation}
\noindent where identities (\ref{a5}) have been taken into account. 

\noindent Let us now prove that $h^i\sim O(\epsilon)$ and $\sum_{i=1}^rJ_ih^i\sim O(\epsilon^2)$ where $\epsilon\equiv (\varrho-l)/l\ll 1$. The definition of $h^i$ is reported in Eq.~(\ref{nlcr21})
\begin{equation}\label{a7}
h^i(X)\equiv \sum_{j,p,q=1}^r\bigl( {\dot\varrho}^{-1}M^{ij}J_j\bigr)\!_{,p}\ {\dot\varrho}^{-1}M^{pq}J_q-{\dot\varrho}^{-2}
\sum_{j,p,p',q=1}^r\Gamma^i_{jp}M^{jp'}M^{pq}J_{p'}J_q
\end{equation}
\noindent Taking into account definition (\ref{a3}) and expansions (\ref{nlcr12}), we have
\begin{equation}\label{a8}
\left\{ \begin{array}{ll}
g_{ij}=g_{ij}^{(0)}+ g_{ij}^{(2)}(\varrho-l)^2+O(\epsilon^4)  &\\
\sum_{p=1}^rg_{ij,p}\frac{dX^p}{d\varrho}= \frac{dg_{ij}}{d\varrho}=2g_{ij}^{(2)}(\varrho-l)+42g_{ij}^{(4)}(\varrho-l)^3+O(\epsilon^5) &\\
f_{ij}=f_{ij}^{(0)}+ f_{ij}^{(2)}(\varrho-l)^2+O(\epsilon^4) & \\
\sum_{p=1}^rf_{ij,p}\frac{dX^p}{d\varrho}= \frac{df_{ij}}{d\varrho}=2f_{ij}^{(2)}(\varrho-l)+42f_{ij}^{(4)}(\varrho-l)^3+O(\epsilon^5) &
\end{array}
\right.
\end{equation}
\noindent Considering that $dX^i/d\varrho$ expands as 
\begin{equation}\label{exp}
\frac{dX^i}{d\varrho}=X^i_{(1)}+3X^i_{(3)}(\varrho-l)^2+O(\epsilon^4)
\end{equation}
\noindent the terms of the second expansion in Eqs.~(\ref{a8}) are balanced if we have 
\begin{equation}\label{a10}
\sum_{i,j=1}^rg_{ij,p}\frac{dX^i}{d\varrho}\frac{dX^j}{d\varrho}=(\varrho-l)\sum_{i,j=1}^rg_{ij,p}^{(1)}X^i_{(1)}X^j_{(1)}+O(\epsilon^3)
\end{equation}
%\noindent Note the following symmetry
%\begin{equation}\label{a11}
%\left\{ \begin{array}{ll}
%(\varrho-l)\rightarrow -(\varrho-l)&\\
%X^\mu_{st.}\rightarrow -X^\mu_{st.}&
%\end{array}
%\right.
%\qquad  \Longrightarrow\qquad
%\left\{ \begin{array}{ll}
%{\check N}^{ij}\rightarrow {\check N}^{ij} & \\
%\sigma\rightarrow \sigma  &
%\end{array}
%\right.
%\end{equation}
\noindent Hence, close to the steady-state, we have
\begin{eqnarray}\label{a12}
&&\sigma=\sigma_{st.}+O(\epsilon) \quad;\quad
\sigma_{st.}\equiv\sum_{i,j=1}^rg_{ij}^{(0)}X^i_{st.}X^j_{st.}
\nonumber\\
&&{\check N}^{ij}=\Bigl[g_{ij}^{(0)}+\frac{1}{\sigma_{st.}}\sum_{p,q=1}^r(f_{ip}^{(0)}g_{jq}^{(0)}+f_{jp}^{(0)}g_{iq}^{(0)})X^p_{st.}X^q_{st.}\Bigr]+O(\epsilon) 
\end{eqnarray}
\noindent By inserting expansions (\ref{nlcr12}), (\ref{a8}), (\ref{a10}) and (\ref{a12}) into Eq.~(\ref{a1}), we obtain 
\begin{equation}\label{a13}
\Gamma^i_{jp}= \Gamma^i_{jp(1)} (\varrho-l)+O(\epsilon^3)
\end{equation}
\noindent where $\Gamma^i_{jp(1)}$ is the expression (\ref{a1}) evaluated at the lowest order. Noting that
\begin{equation}\label{a14}
 \sum_{j,p,q=1}^r\bigl( {\dot\varrho}^{-1}M^{ij}J_j\bigr)\!_{,p}\ {\dot\varrho}^{-1}M^{pq}J_q
 =-6X^i_{(1)}(\varrho-l)+O(\epsilon^3)
 \end{equation}
 \noindent From expansions (\ref{nlcr12}), (\ref{a13}) and (\ref{a14}), we finally get
 \begin{equation}\label{a15}
 h^i=- \Bigl[6X^i_{(1)}+\sum_{j,p=1}^r\Gamma^i_{jp(1)}X^j_{(1)}X^p_{(1)}\Bigr](\varrho-l)+O(\epsilon^3)
 \end{equation}
 \noindent Thus, at the lowest order 
 \begin{equation}\label{a15a}
 h^i\sim O(\epsilon)
 \end{equation}
 \noindent Multiplying Eq.~(\ref{nlcr21}) by $J_i$ and summing over the index "$i$", we obtain 
\begin{eqnarray}\label{a16}
\sum_{i=1}^rJ_i h^i= &&\frac{d}{d\varrho}\Bigl[{\dot\sigma}^{-1}\!\sum_{i,j=1}^rM_{(-1)}^{ij}J_i^{(1)}J_j^{(1)}(\varrho-l)+O(\epsilon^3)\Bigr]\\
&&+{\dot\sigma}^{-2}\!\!\!\!\sum_{i,j,p,q=1}^r\tau_{ij}(X_{st.})M_{(-1)}^{ip}M_{(-1)}^{jq}
J_p^{(1)}J_q^{(1)}+O(\epsilon^2)\nonumber
 \end{eqnarray}
 \noindent where Eq.~(\ref{a6}) has been taken into account. Eq.~(\ref{a16}) may also be rewritten as 
\begin{eqnarray}\label{a17}
\sum_{i=1}^rJ_i h^i\!\!\!\!&&=- \frac{d}{d\varrho}\Bigl[\sum_{i=1}^rJ_i^{(1)}X^i_{(1)}(\varrho-l)+O(\epsilon^3)\Bigr]+\sum_{i,j=1}^r\tau_{ij}(X_{st.})X^i_{(1)}X^j_{(1)}+O(\epsilon^2)\\
\!\!\!\!&&=-\sum_{i=1}^rJ_i^{(1)}X_{(1)}^i+\sum_{i,j=1}^r\tau_{ij}(X_{st.})X^i_{(1)}X^j_{(1)}+O(\epsilon^2)
\nonumber
\end{eqnarray}
\noindent However, from Eq.~(\ref{nlcr7}) and Eq.~(\ref{nlcr13}), and taking into account that $\sum_{i=1}^rJ_i(X_{st.})X_{(1)}^i=0$, we have 
 \begin{equation}\label{a18}
\sum_{i=1}^rJ_i^{(1)}X_{(1)}^i=\sum_{i,j=1}^r\tau_{ij}(X_{st.})X^i_{(1)}X^j_{(1)}+O(\epsilon^2)
\end{equation}
\noindent Hence, at the lowest order, we finally get 
\begin{equation}\label{a19}
\sum_{i=1}^rJ_i h^i\sim O(\epsilon^2)
\end{equation}

%%%%%%%%%%%%%%%%%%%%%END_TEXT_PAPER%%%%%%%%%%%%%%%%%%%%

%%%%%%%%%%%%%%%%%%%%%BIBLIOGRAPHY%%%%%%%%%%%%%%%%%%%%%%

%%%%%%%%%%%%%%%%%%%%%END_BIBLIOGRAPHY%%%%%%%%%%%%%%%%%%%


\begin{thebibliography}{alpha}

\bibitem{sonninog} G. Sonnino, {\it Nonlinear Closure Relation Theory for Transport Processes in Non-Equilib. Systems}, Phys. Rev. E, {\textbf 79}, 051126 (2009). 

\bibitem{sonnino3} G. Sonnino and P. Peeters {\it Physics of Plasmas} {\bf 15}, 062309/1-062309/23 (2008).

\bibitem{onsager} L. Onsager {\it Phys. Rev.}, {\bf 37}, 405 (1931).

\noindent L. Onsager {\it Phys. Rev.}, {\bf 38}, 2265 (1931).

\bibitem{vidal} C. Vidal, G. Dewel and P. Borckmans {\it Au-del{\`a} de l'{\'e}quilibre}, (Hermann {\'E}diteurs Des Sciences et des Arts, Paris, 1994) 341. 

\bibitem{dedonder} Th. De Donder {\it Bull. classe sci. Acad. roy. Belg.}, {\bf 23}, 244 (1937).

\bibitem{prigogine_cov} I. Prigogine 1954 {\it Thermodynamics of Irreversible processes}, (John Wiley \& Sons) 42. 

\noindent I. Prigogine and R. Hansen  {\it Bull. classe sci. Acad. roy. Belg.}, {\bf 28}, 301 (1942).

\noindent I. Prigogine {\it Bull. classe sci. Acad. roy. Belg.}, {\bf 32}, 30 (1946).

\noindent J. Meixner {\it Ann. Physik} [5], {\bf 41}, 409 (1942).

\noindent J. Meixner {\it Ann. Physik} [5], {\bf 43}, 244 (1943).

\bibitem{dedonder1} Th. De Donder {\it L'Affinit{\'e}}, (Gauthier-Villars \& ${\rm C}^{\rm le}$ Editeurs, Paris, 1936) 32

\noindent Th. De Donder {\it L'Affinit{\'e} (Seconde Partie)}, (Gauthier-Villars \& ${\rm C}^{\rm le}$ Editeurs, Paris, 1931) 16. 

\bibitem{degroot} S.R. De Groot and P. Mazur {\it Non-Equilibrium Thermodynamics}, (Dover Publications, Inc., New York, 1984) 459. 

\bibitem{prigogine1} P. Glansdorff and I. Prigogine {\it Physica}, {\bf 20}, 773 (1954).

\noindent P. Glansdorff and I. Prigogine {\it Thermodynamic Theory of Structure, Stability and Fluctuations}, (John Wiley \& Sons, London, New york 1971), 106. 

\bibitem{haken} H. Haken {\it Synergetics. Introduction and Advanced Topics}, (Springer, Berlin, Heidelberg, New York Tokio, 1976) 202, 204.

\noindent  M. Malek Mansour, J. Dethier and  F. Baras {\it J. Chem. Phys.}, {\bf 144}, 9265 (2001).

\noindent H. Haken {\it Physica D} {\bf 97}, 95 (1996).

\bibitem{balescu2} R. Balescu {\it Transport Processes in Plasmas. Vol 2. Neoclassical Transport}, (Elsevier Science Publishers B.V., Amsterdam, North-Holland, 1988) 423.

\bibitem{davidson} R.C. Davidson, {\it Methods in Nonlinear Plasma Theory}, (Academic Press, New York, 1972). 

\bibitem {eisenhart} L.P. Eisenhart {\it Non-Riemannian Geometry}, American Mathematical Society, Colloquium Publications (Providence, Rhode Island), Vol. 8, (1927). 

\bibitem{sonnino} G. Sonnino {\it Il Nuovo Cimento}, {\bf 115 B}, 1057 (2000). 

\noindent G. Sonnino {\it Thermodynamic Field Theory (An Approach to Thermodynamics of Irreversible Processes)} proceedings of the {\it 9th International Workshop on Instabilities and Nonequilibrium Structures}, Vi$\tilde{\mathrm n}$a del Mar (Chile), (Kluwer Academic Publishers, 2001), 291. 

\noindent G. Sonnino {\it A Field Theory Approach to Thermodynamics of Irreversible Processes}, (Th{\` e}se d'Habilitation {\`a} Diriger des Recherches - H.D.R.) 
- Institut Non Lin{\`e}aire de Nice (I.N.L.N.) (Nice, France, 2002). 

\noindent G. Sonnino {\it Nuovo Cimento}, {\bf 118 B}, 1115 (2003). 

\end{thebibliography}
\end{document}